\newcommand{\mua}{{\mu_{a}}}
\newcommand{\mus}{{\mu_{s}}}

\renewcommand{\vec}[1]{\boldsymbol{#1}}
\newcommand{\svec}{\hat{\vec{s}}}
\newcommand{\xvec}{\vec{x}}
\newcommand{\rpos}{\mathbf{r}}

\newcommand{\text}[1]{\textrm{#1}}
\newcommand{\grad}[2]{\frac{\partial #1}{\partial #2}}
\newcommand{\adj}{^{\ast}}

\newcommand{\adomain}{\mathcal{S}^{n-1}}
\newcommand{\fluence}{\Phi}
\newcommand{\radiance}{\phi}

\documentclass[12pt]{spieman}
\usepackage{amsmath}
\usepackage{amssymb}
\expandafter\let\csname equation*\endcsname\relax
\expandafter\let\csname endequation*\endcsname\relax
\usepackage{amsmath}
\usepackage{comment}
\usepackage{comment}
\usepackage[]{graphicx}
\usepackage{epstopdf}
\usepackage{float}
\usepackage{color}

\title{Quantitative photoacoustic tomography using forward and adjoint Monte Carlo models of radiance}
\author{Roman Hochuli, Samuel Powell, Simon Arridge, Ben Cox}

\begin{document}
\maketitle

\begin{abstract}
Forward and adjoint Monte Carlo (MC) models of radiance are proposed for use in model-based quantitative photoacoustic tomography. A 2D radiance MC model using a harmonic angular basis is introduced and validated against analytic solutions for the radiance in heterogeneous media. A gradient-based optimisation scheme is then used to recover 2D absorption and scattering coefficients distributions from simulated photoacoustic measurements. It is shown that the functional gradients, which are a challenge to compute efficiently using MC models, can be calculated directly from the coefficients of the harmonic angular basis used in the forward and adjoint models. This work establishes a framework for transport-based quantitative photoacoustic tomography that can fully exploit emerging highly parallel computing architectures.
\end{abstract}

\section{Introduction}\label{sec:introduction}
Quantitative photoacoustic tomography is concerned with recovering quantitatively accurate estimates of chromophore concentration distributions, or related quantities such as optical coefficients or blood oxygenation, from photoacoustic images \cite{Cox2012}. The source of contrast in photoacoustic tomography (PAT) is optical absorption, which is directly related to the tissue constituents. By obtaining PAT images at multiple optical wavelengths, it may be possible to recover chemically specific information about the tissue. However, such a spectroscopic use of PAT images must consider the effect of the spatially and spectrally varying light fluence distribution. As a photoacoustic image is the product of the optical absorption coefficient distribution, which carries information about the tissue constituents, and the optical fluence, which only acts to distort that information, the challenge in quantitative photoacoustic imaging is to remove the effect of the light fluence.

A common approach is to use a model of the unknown fluence and use it to extract the desired optical properties from the measured data. This has been done analytically \cite{Bal2010a,Bal2010b,Ren2013,Ammari2010a} or numerically \cite{Zemp2010b,Harrison2013}, often within a minimisation framework \cite{Laufer2010a,Cox2009,Tarvainen2013,Malone2015,Saratoon2013,Pulkkinen2014,Ding2015a,Yao2009a}. The majority of this literature uses the diffusion approximation to the radiative transfer equation to model the light distribution, which is accurate in highly scattering media and away from boundaries or sources \cite{Arridge1999a}. In PAT, the region of interest often lies close to the tissue surface where the diffusion approximation is not accurate. The radiative transfer equation (RTE), on the other hand, is widely considered to be an accurate model of light transport so long as coherent effects are negligible, which is the case here. Finite element discretisations of the RTE have been developed \cite{Tarvainen2006a,SuryaMohan2011} and proposed for quantitative PAT reconstructions \cite{Saratoon2013,Yao2009a}, but due to the need to discretise in angle as well as space they quickly become computationally intensive and their applicability is limited to small and medium scale problems. An alternative is Monte Carlo (MC) modelling \cite{Wang1995a,Boas2002,Fang2009,Powell2012}, which is a stochastic technique for modelling light transport that converges to the solution to the RTE. The significant advantage of the MC approach is that it is highly parallelisable so scales well to the large-scale inversions that will be encountered in practice.

Monte Carlo models of light transport are popular in biomedical optics and have predominantly been applied in the planning of experimental measurements \cite{Beard1997,Antonelli2011,Leung2013a} and in dosimetric studies for a range of light based therapies \cite{Grosges2011,Manuchehrabadi2013,Cassidy2015}. Many of the applications are summarised by Zhu et al. \cite{Zhu2013a}. One early MC model of light transport, MCML \cite{Wang1995a}, computes the fluence in 3D slab geometry. This model was later extended to simulate spherical inclusions in the tissue\cite{Periyasamy2013}, and later to spheroidal and cylindrical \cite{Periyasamy2014} inclusions. MC modelling in 3D heterogeneous media has been shown both for voxelised media \cite{Boas2002}, which was later GPU-accelerated \cite{Fang2009}, and using a mesh-based geometry \cite{Fang2010,Powell2012}. Although the RTE is an equation for the radiance, which is a function of angle at every point, the quantity usually calculated by MC models is the fluence rate, which is the radiance integrated over all angles. The reasons are practical: most measurable quantities are related to the fluence rate rather than the radiance, storing just the integrated quantity saves on computational memory, and the estimates for the fluence rate will converge sooner than the underlying estimates for the radiance. In photoacoustics, the measurable signal is related to the fluence (the time-integrated fluence rate) so current MC models can be used in the simulation of photoacoustic signals. However, as will be discussed in Section 4, the full angle-dependent radiance is required when tackling the inverse problem of estimating the optical coefficients, specifically the optical scattering.

In this paper, Section \ref{sec:QPAT} introduces the inverse problem of quantitative PAT. Sections \ref{sec:light_transport} and \ref{sec:adjoint} present forward and adjoint Monte Carlo models of the radiance employing a harmonic angular basis. In Section \ref{sec:gradients} it is shown that this choice of basis allows the functional gradients for the inverse problem to be calculated straightforwardly. Inversions for absorption and scattering coefficient distributions are given in Section \ref{sec:examples}.

\section{Quantitative Photoacoustic Tomography}\label{sec:QPAT}
The inverse problem in QPAT can be stated as the minimisation
\begin{align}
\underset{\mu_a(\xvec),\mu_s(\xvec)}{\text{argmin}} \,\, \epsilon(\mu_a(\xvec),\mu_s(\xvec))
\end{align}
where the error functional is given by
\begin{align}
\epsilon = \frac{1}{2}\int_{\Omega}\left(H^{meas}(\xvec)-H(\xvec; ~\mua,\mus)\right)^{2}d\xvec. \label{error_function3}
\end{align}
$H=\mua(\xvec,\lambda)\Phi(\xvec,\lambda;~\mua,\mus,g)$ is the absorbed energy density and is the `data' for this problem. It is related to the photoacoustic image by the Gr\"uneisen parameter, which here is set to 1.  Additional regularisation terms or terms reflecting prior knowledge may also be added to $\epsilon$. Gradient-based approaches to solving this problem require estimates of the gradients of the error functional with respect to the parameters of interest. Saratoon et al. \cite{Saratoon2013} gives expressions for these gradients in terms of the forward and adjoint fields, $\radiance$ and $\radiance\adj$:
\begin{equation}
\grad{\epsilon}{\mua} = - \fluence(H^{meas}-H) + \int_{\adomain}\radiance\adj(\svec)\radiance(\svec) d\svec,
\label{absorption_gradient_main}
\end{equation}
and
\begin{equation}
\grad{\epsilon}{\mus} = \int_{\adomain} \radiance\adj(\svec)\radiance(\svec) d\svec - \int_{\adomain}\int_{\adomain} \radiance\adj(\svec)P(\svec,\svec')\radiance(\svec') d\svec'd\svec.
\label{scattering_gradient_main}
\end{equation}
Monte Carlo models to calculate the radiance $\radiance(\svec)$ and adjoint radiance $\radiance\adj(\svec)$ are given in the following two sections.

\section{Monte Carlo Modelling of Light Transport}\label{sec:light_transport}
In PAT, the optical and acoustic propagation times are so different that the optical propagation can be considered instantaneous and the time-dependence of the light transport can be neglected. The time-independent radiative transfer equation (RTE) is given by
\begin{equation}
(\svec\cdot\nabla + \mua(\xvec)+\mus(\xvec))\phi(\xvec,\svec)-\mus(\xvec)\int_{\adomain}P_{\theta}(\svec,\svec')\phi(\xvec,\svec')d\svec' = q(\xvec,\svec),
\label{RTE}
\end{equation}
where $\phi$ is the radiance, $\mua$ and $\mus$ are the absorption and scattering coefficients, respectively, $\xvec$ is position, $\svec'$ and $\svec$ are the original and scattered propagation directions, $P_{\theta}(\svec,\svec')$ is the scattering phase function, $q(\xvec,\svec)$ is a source term and $\adomain$ is used to indicate integration over angle in $n-1$ dimensions. To obtain approximations to the solutions to this equation, various flavours of MC have been proposed \cite{Sassaroli2012}. The approach used here begins with launching a packet of energy, referred to herein as a `photon', from a given position $\xvec$ in an initial direction $\svec$. After travelling a distance $s=\mathcal{U}([0,1]) / \mus$ (using the convention $\mathbf{s} = \left|\mathbf{s}\right|\svec$), where $\mathcal{U}([0,1])$ is a real uniform random variable on $[0,1]$, a fraction of photon's `weight' $W(1-\exp(-\mua s))$ is deposited in current voxel, where $W$ is the current weight (or energy) of the photon packet. The photon weight is updated: $W \leftarrow W\exp(-\mua s)$. Scattering into a new direction $\svec'$ in 2D involves sampling the scattering phase function, which describes the probability of a photon scattering from direction $\svec'$ into direction $\svec$. The phase function used here was the 2D Henyey-Greenstein phase function, commonly used in biomedical optics \cite{Heino2003,Star1988},
\begin{equation}
P(\svec,\svec') =\frac{1}{2\pi}\frac{1-g^{2}}{(1+g^{2}-2g(\svec\cdot\svec'))}.
\label{scattering2D}
\end{equation}
The parameter, $g$, a property of the medium, is known as the anisotropy factor. Sampling this equation for the scattering angle, $\theta = \arccos(\svec\cdot\svec')$, by solving for $\theta$ in the cumulative integral over angle yields
\begin{equation}
\theta = 2\arctan\left(\frac{1-g}{1+g}\tan(\pi\mathcal{U}([0,1]))\right).
\label{scattering_angle}
\end{equation}
A new step length, $s$, is sampled and this process is repeated until the photon weight falls below some threshold value. By carrying out the above computation for many photons and adding the voxel weights will, for a sufficient number of photons, converge on a solution to the RTE.

By calculating photon paths through the medium, the MC models presented in the literature \cite{Wang1995a,Boas2002,Fang2009,Fang2010} do in fact simulate the radiance, but this typically integrated over angle upon deposition of the weights in the voxels. In order to simulate the radiance, a method of depositing the weight in the voxels without losing the angular information is required.

\subsection{Monte Carlo modelling of the radiance}\label{sec:radiance}
In order to compute the radiance using a MC model, angular as well as spatial discretisation is required. One approach is to use discrete ordinates, whereby the unit circle is divided equally into sectors and the weight deposited in a voxel is also assigned to the relevant angular sector. The memory required will scale linearly with the number of sectors, and will slow convergence of the radiance estimate, compared with the fluence estimate, by a factor inversely related to the number of sectors. Here, a harmonic angular basis was used because a sufficiently diffuse field is dense in such a basis, meaning the field can be represented using relatively few orders. Less memory will therefore be required.

In 2D, the expansion for the radiance in a Fourier basis is \cite{Hochuli2015a}:
\begin{equation}
\phi(\mathbf{x},\theta)=\frac{1}{2\pi}a_{0}(\mathbf{x})+\frac{1}{\pi}\sum_{n=1}^{N=\infty}a_{n}(\mathbf{x})\cos(n\theta)+\frac{1}{\pi}\sum_{n=1}^{N=\infty}b_{n}(\mathbf{x})\sin(n\theta),
\label{radiance_Fourier}
\end{equation}
where $a_n$ and $b_n$ are the coefficients associated with each harmonic and $\theta\in[-\pi,\pi]$ and is the angle of the photon direction $\svec$ relative to the z-direction (i.e. $\theta=\arccos(\svec)$). (The equivalent expansion in 3D would be into spherical harmonics \cite{Hochuli2016}.)
For a given voxel, the weight is deposited into the relevant Fourier coefficients according to
\begin{align}\label{Fourier_terms}
& a_{0} = \sum_{n_{p}=1}^{N_{p}}dW_{n_{p}}\int_{\mathcal{S}^{1}}\delta(\theta'-\theta_{n_{p}})d\theta' = \sum_{n_{p}=1}^{N_{p}}dW_{n_{p}} \\
& a_{n}= \sum_{n_{p}=1}^{N_{p}}dW_{n_{p}}\int_{\mathcal{S}^{1}}\delta(\theta'-\theta_{n_{p}})\cos(n\theta)d\theta' = \sum_{n_{p}=1}^{N_{p}}dW_{n_{p}}\cos(n\theta_{n_{p}}) \\
& b_{n} = \sum_{n_{p}=1}^{N_{p}}dW_{n_{p}}\int_{\mathcal{S}^{1}}\delta(\theta'-\theta_{n_{p}})\sin(n\theta)d\theta' = \sum_{n_{p}=1}^{N_{p}}dW_{n_{p}}\sin(n\theta_{n_{p}}),
\end{align}
where $dW_{n_{p}}$ is the weight deposited by the $n_{p}^{\text{th}}$ photon traversing the $i$\textsuperscript{th} voxel. The algorithm was implemented in the Julia programming language \cite{Bezanson2014}.

\subsection{Validation of the forward model}\label{sec:validation}
\begin{figure}[H]
\centering
\includegraphics[width=1.0\textwidth]{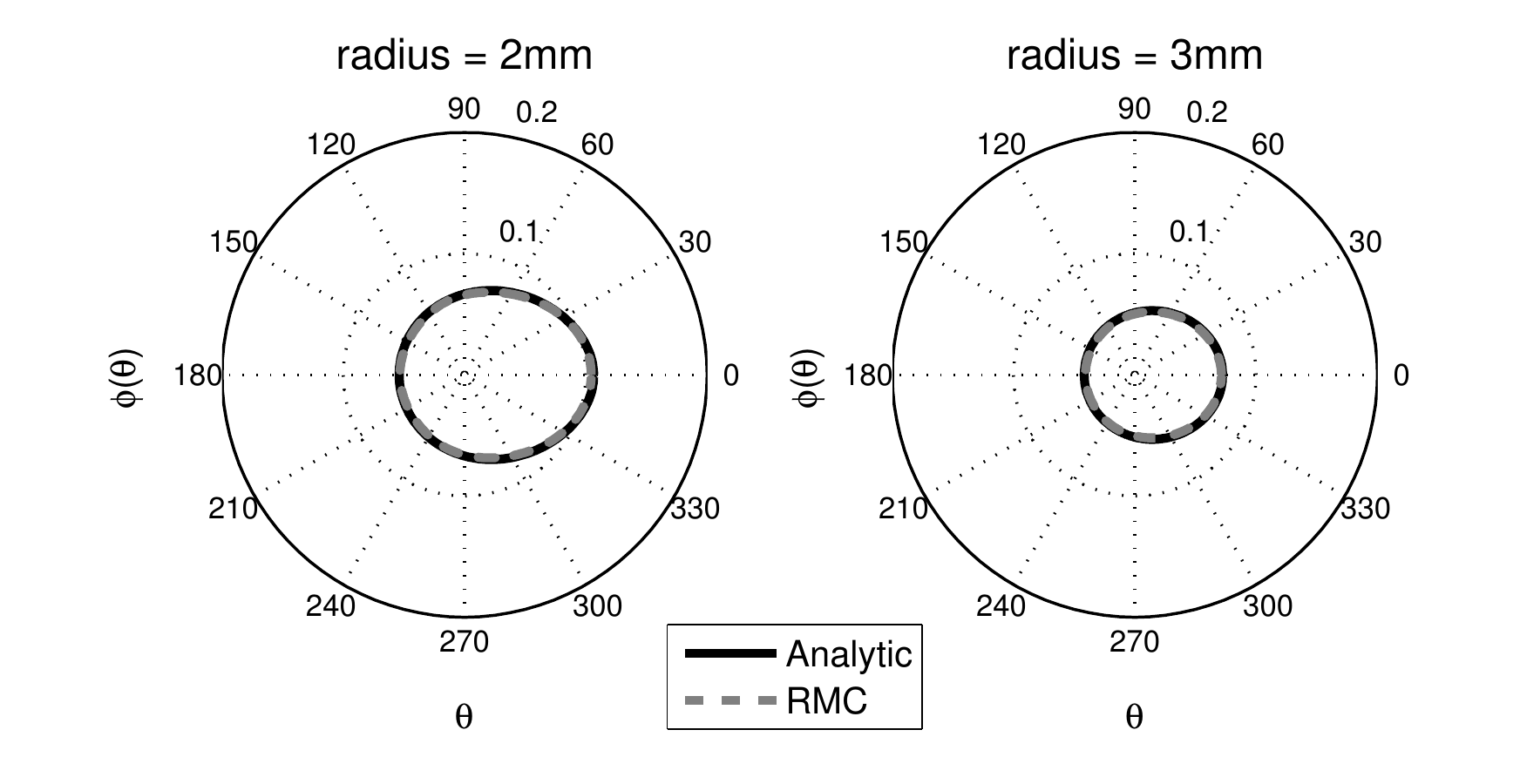}
\caption{Polar plots of the angle-resolved radiance due to an isotropic point source in a homogeneous domain with $\mua=0.01$mm\textsuperscript{-1}, $\mus=10$mm\textsuperscript{-1} and $g=0.9$. Results from an analytic method (infinite domain) and RMC simulations (15mm$\times$15mm square domain) shown.}
\label{fig:RMC_validation}
\end{figure}

Analytical solutions to the RTE are available for the fluence for a range of geometries and source types \cite{Boas2002,VanGemert1987,Jha2012}, however there are few analytical solutions for the radiance, particularly in 2D. The RMC model was compared to one such analytic solution for an infinite, homogeneous 2D domain illuminated by an isotropic point source \cite{Liemert2012b,Liemert2011,Liemert2013}. An isotropic point source was placed at the centre of a domain of size 15mm$\times$15mm, large compared to the transport mean free path in order to approximate an infinite domain. The absorption and scattering coefficients were 0.01mm\textsuperscript{-1} and 10mm\textsuperscript{-1} respectively and the Henyey-Greenstein phase function \cite{Henyey1940} was used used with $g$ set to 0.9. The pixel size was 0.05mm $\times$ 0.05mm, and 5 Fourier harmonics were used. Fig.\ \ref{fig:RMC_validation} shows the good agreement between the analytical and RMC modelled radiance at radial distances of 2mm and 3mm from the source along the horizontal axis.

\section{Adjoint Monte Carlo model}\label{sec:adjoint}
The adjoint equation to the RTE is given by
\begin{equation}
(-\svec \cdot \nabla+\mua(\xvec)+\mus(\xvec))\phi^{*}(\xvec,\svec)=\mus(\xvec)\int_{\mathcal{S}^{n-1}}P(\svec',\svec)\phi^{*}(\xvec,\svec')d\svec'+q^{*}(\xvec,\svec),
\label{adjoint_RTE1}
\end{equation}
where $\phi^{*}$ is the adjoint radiance and $q^{*}$ is the adjoint source. This was implemented numerically using the same MC scheme as for the forward RMC model (Section \ref{sec:radiance}). The principle difference is that the light sources $q$ typically used in PAT are restricted to the boundary, but the adjoint source $q^{*}$ will not be, as a consequence of the fact that the `data' in QPAT - the photoacoustic images - is volumetric.

\subsection{Validation of the adjoint model}
The adjoint model was validated by checking it satisfied the condition:
\begin{align}
\langle\mathcal{L}\mathbf{a},\mathbf{b}\rangle=\langle\mathbf{a},\mathcal{L}^{*}\mathbf{b}\rangle,
\label{eq:adj_definition}
\end{align}
where $\mathcal{L}$ and $\mathcal{L}^{*}$ are the operators corresponding to the forward and adjoint RMC models, and $\mathbf{a}$ and $\mathbf{b}$ are the angle and position dependent source and detector. Three cases were tested:
\begin{eqnarray}
\text{Case 1: } & \mathbf{a}_{1}=\delta(\rpos-\rpos_{s})/2\pi, \,\, & \mathbf{b}_{1}=\delta(\rpos-\rpos_{d}) \label{adj_case1} \\
\text{Case 2: } & \mathbf{a}_{2}=\delta(\rpos-\rpos_{s})/2\pi, \,\, & \mathbf{b}_{2}=\delta(\rpos-\rpos_{d})P_d(\rpos_{d},\svec) \label{adj_case2}\\
\text{Case 3: } & \mathbf{a}_{3}=P_{s}(\rpos), \,\, & \mathbf{b}_{3}=\delta(\rpos-\rpos_{d})P_d(\rpos_{d},\svec), \label{adj_case3}
\end{eqnarray}
where $\rpos_{s}$ and $\rpos_{d}$ are the positions of the source and detector, $P_{d,s}(\rpos,\svec)$ are the spatial and angular sensitivity of the detector and source. Substituting these into \eqref{eq:adj_definition} yields
\begin{align}
\Phi_{1}(\rpos_{d}) &= \Phi^{*}_{1}(\rpos_{s}), \label{eq:case1_2} \\
\int_{2\pi}\phi_{2}(\rpos_{d},\svec)P_{d}(\rpos_{d},\svec)d\svec &= \Phi^{*}_{2}(\rpos_{s}), \label{eq:case2_2} \\
2\pi\int_{\Omega}\Phi_{3}(\rpos)P_{d}(\rpos)d\rpos &= \int_{2\pi}\phi^{*}_{3}(\rpos_{d})P_{s}(\svec)d\svec. \label{eq:case3_2}
\end{align}
where $\phi_{1,2,3}$ and $\phi^*_{1,2,3}$ are the forward and adjoint radiances from computing $\mathcal{L}\mathbf{a}_{1,2,3}$ and $\mathcal{L}^{*}\mathbf{b}_{1,2,3}$, respectively. $\Phi$ is the fluence, or angle-integrated radiance. It can be seen from \eqref{eq:case1_2} that the case where a pair of isotropic $\delta$-functions are used for $\mathbf{a}_{1}$ and $\mathbf{b}_{1}$, that we expect the resulting fluence values at their respective positions, $\Phi_{1}(\rpos_{d})$ and $\Phi^{*}_{1}(\rpos_{s})$, to be equal. This is an intuitive result given the reciprocity of the RTE and the angular indpendence of the source-detector combination.

\begin{figure}[H]
\centering
\includegraphics[width=0.6\textwidth]{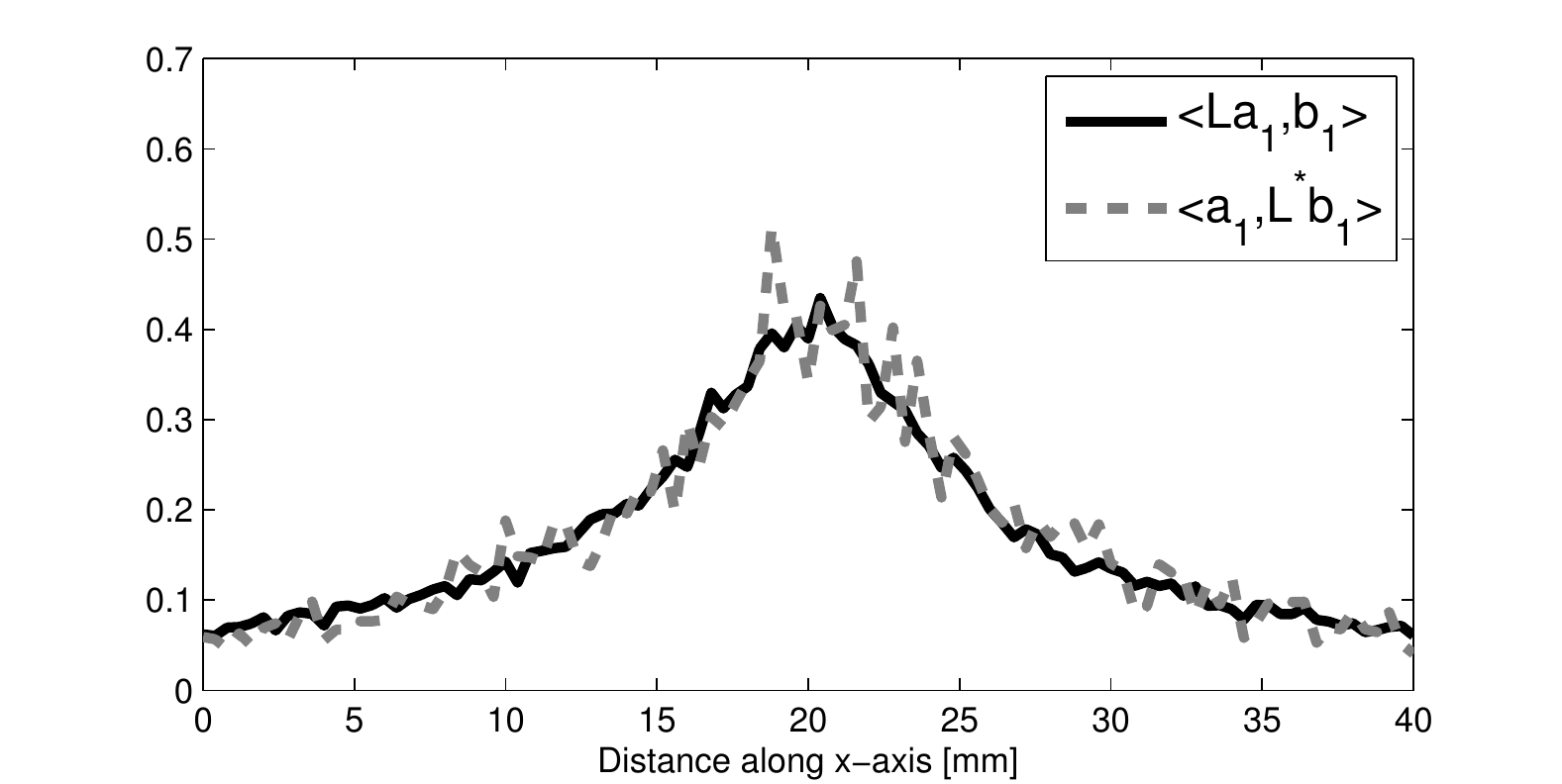}
\caption[Plots showing evaluation of the definition of the adjoint model using isotropic point sources/detectors]{Plot of $\langle\mathcal{L}\mathbf{a}_1,\mathbf{b}_1\rangle$ and $\langle\mathbf{a}_1,\mathcal{L}^{*}\mathbf{b}_1\rangle$ to validate the adjoint model. $\mathbf{a}_1$ and $\mathbf{b}_1$ were isotropic point sources with $\mathbf{a}_1$ at the centre of the domain and $\mathbf{b}_1$ translated across the domain at y = 23.6mm.}
\label{fig:IsoIso_adjoint_test1}
\end{figure}

\begin{figure}[H]
\centering
\includegraphics[width=1.0\textwidth]{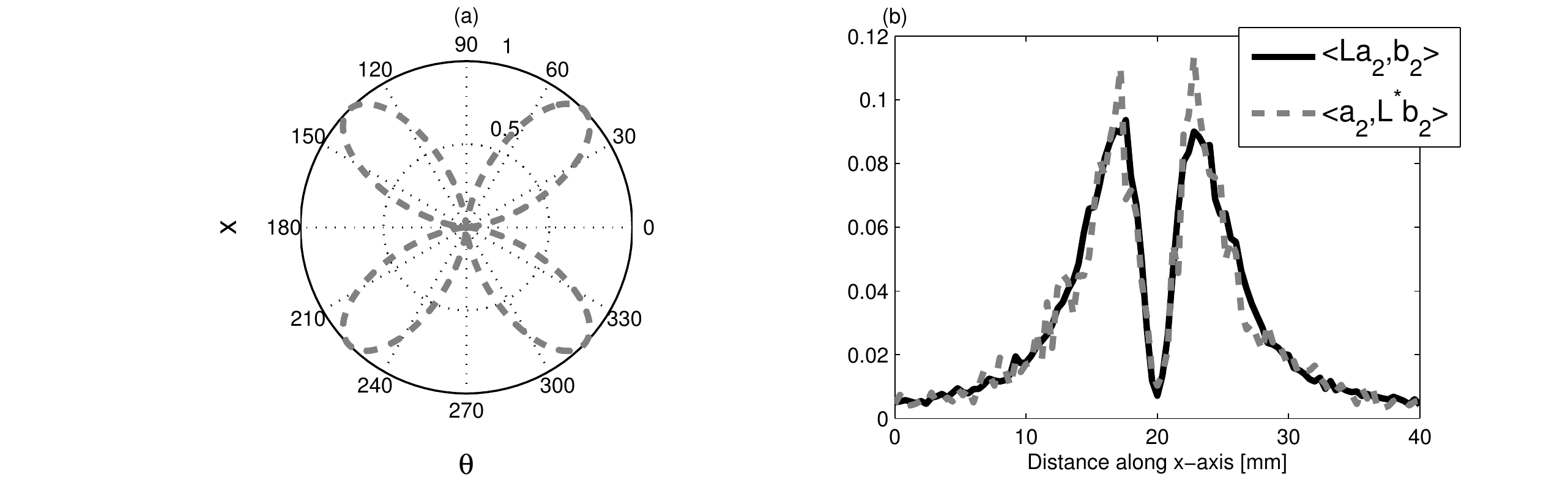}
\caption[Plots showing evaluation of the definition of the adjoint model using an anisotropic point source and an isotropic point detector]{(a) Polar plot of source distribution for $\mathbf{b}_{2}=\delta(\rpos-\rpos_{s})\frac{1}{\pi}\sin^{2}(2\theta)$; (b) Plot of $\langle\mathcal{L}\mathbf{a}_2,\mathbf{b}_2\rangle=\langle\mathbf{a}_2,\mathcal{L}^{*}\mathbf{b}_2\rangle$ for validation of adjoint model. Plot was produced with $\mathbf{a}_2$ as an isotropic point source at the centre of the domain. $\mathbf{b}_2$ was translated across the domain along a line at y = 23.6mm.}		\label{fig:IsoAniso_adjoint_test2}
\end{figure}

\begin{figure}[H]
\centering
\includegraphics[width=1.0\textwidth]{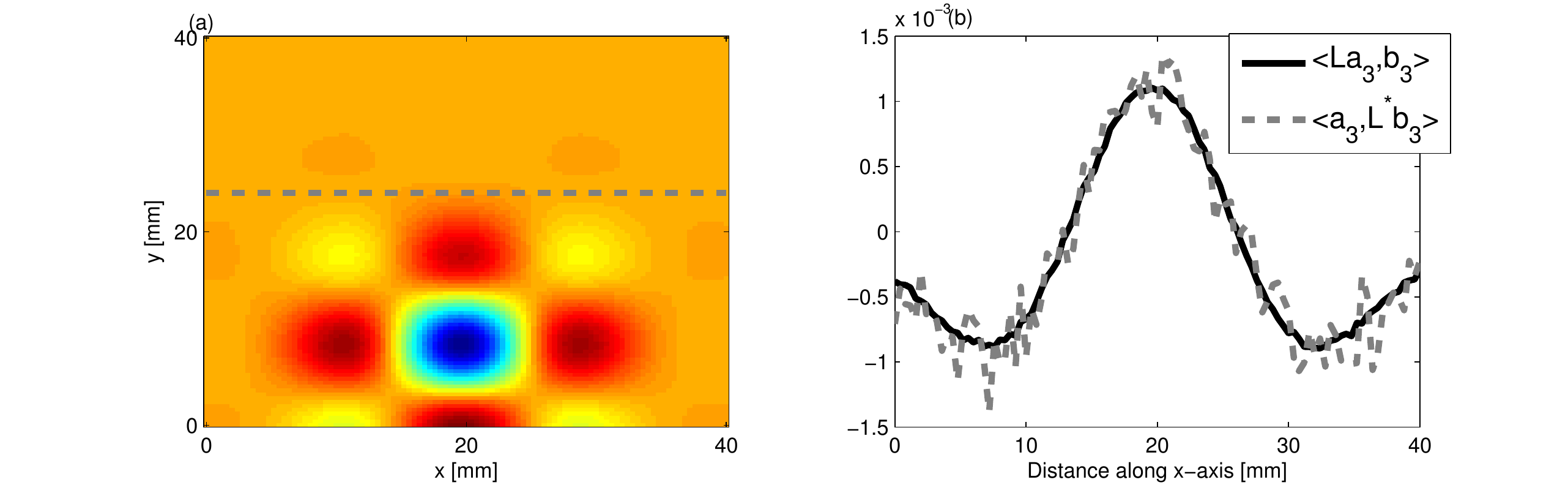}
\caption[Plots showing evaluation of the definition of the adjoint model using an anisotropic point source and a distribution of point detectors]{(a) Isotropic source distribution $\mathbf{a}_{3}=P_{s}(\rpos)$; (b) $\langle\mathcal{L}\mathbf{a}_3,\mathbf{b}_3\rangle$ and $\langle\mathbf{a}_3,\mathcal{L}^{*}\mathbf{b}_3\rangle$ to validate the adjoint model. $\mathbf{b}_3$ was an anisotropic point source emitting light over angle following $\frac{1}{\pi}\sin^{2}(2\theta)$. $\mathbf{b}_3$ was translated along a line across the domain at y = 23.6mm, as shown by the grey line dashed line in (a).}
\label{fig:IsoIso_adjoint_test3}
\end{figure}

Simulations were performed using a 40mm$\times$40mm (101$\times$101 pixel) domain, and 10 Fourier harmonics. Each source distribution emitted 10\textsuperscript{6} photons. $\rpos_{s}$ was set to be the centre of the domain with $\rpos_{d}$ moved along the x-direction across the domain. Comparisons are shown in Fig. \ref{fig:IsoIso_adjoint_test1} for Case 1 with an isotropic source and detector, Fig. \ref{fig:IsoAniso_adjoint_test2} for Case 2 with an isotropic source and and anisotropic detector with $P_d=\delta(\rpos-\rpos_{s})\frac{1}{\pi}\sin^{2}(2\theta)$, and Fig.\ref{fig:IsoIso_adjoint_test3} for Case 3 with the same $P_d$ but with the distributed $P_{s}(\rpos)$ shown in Fig. \ref{fig:IsoIso_adjoint_test3}(a). Good agreement was obtained in all cases, showing the the RMC adjoint model is an accurate representation of the RTE adjoint.

\section{Functional Gradients}\label{sec:gradients}

Both the radiance and the adjoint radiance can be expressed as Fourier series as in \eqref{radiance_Fourier}. By substituting these expressions into Eqs. \ref{absorption_gradient_main} and \ref{scattering_gradient_main} for the functional gradients, simple and easily computed expressions for the gradients can be obtained. The fluence is simply given by the isotropic component of the field $a_0$. The other terms in the expressions for the functional gradients contain integrals of products of the radiance and its adjoint.
If $a_{0}^{*}$, $a_{n}^{*}$ and $b_{n}^{*}$ are the Fourier coefficients of the adjoint radiance, then the gradient with respect to absorption can be written as
\begin{align}\label{grad_2D_abs_integral1}
\frac{\partial\epsilon}{\partial\mua} =- \fluence(H^{meas}-H) + \int_{\mathcal{S}^{1}}\phi(\svec)\phi^{*}(\svec)d\svec \notag \\ = -a_{0}(H^{meas}-\mua a_{0}) +\int_{2\pi}\left[\frac{1}{4\pi^2}a_{0}a_{0}^{*}+\frac{1}{2\pi^2}a_{0}\sum_{m=1}^{\infty}a_{m}^{*}\cos(m\theta')+\frac{1}{2\pi^2}a_{0}\sum_{m=1}^{\infty}a_{m}^{*}\sin(m\theta') \right. \notag \\
\left.+\frac{1}{2\pi^2}a_{0}^{*}\sum_{n=1}^{\infty}a_{n}\cos(n\theta)+\sum_{n=1}^{\infty}\sum_{m=1}^{\infty}a_{n}a_{m}^{*}\cos(n\theta)\cos(m\theta)+\sum_{n=1}^{\infty}\sum_{m=1}^{\infty}a_{n}b_{m}^{*}cos(n\theta)\sin(m\theta)\right. \notag \\
\left.+\frac{1}{2\pi^2}a_{0}^{*}\sum_{n=1}^{\infty}b_{m}\cos(m\theta)+\sum_{n=1}^{\infty}\sum_{m=1}^{\infty}a_{m}^{*}b_{n}\sin(n\theta)cos(m\theta)+\sum_{n}\sum_{m}b_{n}b_{m}^{*}\sin(n\theta)\sin(m\theta)\right]d\theta.
\end{align}
By orthogonality, all terms for which $n \neq m$ integrate to zero and \eqref{grad_2D_abs_integral1} reduces to
\begin{align}\label{grad_2D_abs_integral2}
\frac{\partial\epsilon}{\partial\mua} & = -a_{0}(H^{meas}-\mua a_{0}) + \int_{2\pi}\left[\frac{1}{4\pi^2}a_{0}a_{0}^{*}+\frac{1}{\pi^2}\sum_{n=1}^{\infty}a_{n}a_{n}^{*}\cos^{2}(n\theta)+\frac{1}{\pi^2}\sum_{n=1}^{\infty}b_{n}b_{n}^{*}\sin^{2}(n\theta)\right]d\theta,\\
& = -a_{0}(H^{meas}-\mua a_{0}) + \frac{1}{2\pi}a_{0}a_{0}^{*}+\frac{1}{\pi}\sum_{n=1}^{\infty}a_{n}a_{n}^{*}+\frac{1}{\pi}\sum_{n=1}^{\infty}b_{n}b_{n}^{*}.
\label{grad_2D_abs_integral3}
\end{align}
This expression for the absorption gradient is computationally straightforward to evaluate due to the fact that it requires simply summing over products of Fourier coefficients already loaded in memory.

The second term in \eqref{scattering_gradient_main} is
\begin{equation}
\int_{\adomain}\int_{\adomain} \radiance\adj(\svec)P(\svec,\svec')\radiance(\svec') d\svec'd\svec,
\end{equation}
which contains the phase function given in \eqref{HGFourier} and can be expanded using a Fourier series in powers of $g$ \cite{Heino2003}:
\begin{align}
P(\svec\cdot\svec'; ~g) = \frac{1}{2\pi}+\frac{1}{\pi}\sum_{l=1}^{\infty}g^{l}\cos(l\Delta\theta), \label{HGFourier}
\end{align}
where $\Delta\theta=\arccos(\svec\cdot\svec')$.
Thus we can write,
\begin{align}\label{grad_2D_scat_integral1}
\int_{2\pi}\int_{2\pi}\phi(\svec')P_{\theta}(\svec,\svec')\phi^{*}(\svec)d\svec d\svec'=
\int_{2\pi}\int_{2\pi} \left[\frac{1}{2\pi}a_{0}+\frac{1}{\pi}\sum_{n=1}^{\infty}a_{n}\cos(n\theta')+\frac{1}{\pi}\sum_{n=1}^{\infty}b_{n}\sin(n\theta')\right] \notag \\
\left[\frac{1}{2\pi}+\frac{1}{\pi}\sum_{l=0}^{\infty}g^{l}\cos(l(\theta-\theta'))\right] \notag \\      \left[\frac{1}{2\pi}a_{0}^{*}+\frac{1}{\pi}\sum_{m=1}^{\infty}a_{m}^{*}\cos(m\theta)+\frac{1}{\pi}\sum_{m=1}^{\infty}b_{m}^{*}\sin(m\theta)\right] d\theta d\theta',
\end{align}
where $\theta$ and $\theta'$ are the angles between the z-axis and $\svec$ and $\svec'$, respectively. As such, the scattering angle between the previous direction $\svec'$ into the new direction $\svec$ is given by $(\theta-\theta')$.
It is possible to expand $\cos(l(\theta-\theta'))$ as $\cos(l\theta)\cos(l\theta')+\sin(l\theta)\sin(l\theta')$ which in turn allows us to employ orthogonality relationships to simplify the above integrals and write
\begin{align}\label{grad_2D_scat_integral2}
\int_{\mathcal{S}^{1}}\int_{\mathcal{S}^{1}}\phi(\svec')P_{\theta}(\svec,\svec')\phi^{*}(\svec)d\svec d\svec'=
\frac{1}{2\pi}a_{0}a_{0}^{*}+\frac{1}{\pi}\sum_{n=1}^{\infty}a_{n}a_{n}^{*}g^{n}+\frac{1}{\pi}\sum_{n=1}^{\infty}b_{n}b_{n}^{*}g^{n},
\end{align}
Substituting this expression into \eqref{scattering_gradient_main}, we can write the full expression for the functional gradient with respect to the scattering coefficient:
\begin{align}
\frac{\partial\epsilon}{\partial\mu_{s}} &= \frac{1}{2\pi}a_{0}a_{0}^{*}+\frac{1}{\pi}\sum_{n=1}^{\infty}a_{n}a_{n}^{*}+\frac{1}{\pi}\sum_{n=1}^{\infty}b_{n}b_{n}^{*} -\frac{1}{2\pi}a_{0}a_{0}^{*}+\frac{1}{\pi}\sum_{n=1}^{\infty}a_{n}a_{n}^{*}g^{n}+\frac{1}{\pi}\sum_{n=1}^{\infty}b_{n}b_{n}^{*}g^{n} \label{scattering_gradient2} \\
&= \frac{1}{\pi}\sum_{n=1}^{\infty}\left[a_{n}a_{n}^{*}+b_{n}b_{n}^{*}\right]\left(1-g^{n}\right). \label{scattering_gradient3}
\end{align}
The ability to calculate these gradients is the first step to finding a computationally efficient way to solve the full QPAT inversion using a Monte Carlo model of light transport.

\section{Inversions for Absorption and Scattering}\label{sec:examples}
The forward and inverse MC models of radiance described above were used with a gradient-descent (GD) scheme to estimate $\mua(\xvec)$ and $\mus(\xvec)$ from simulated PAT images by minimising the error functional in \eqref{error_function3}.  As the adjoint source, $q^{*}(\xvec,\svec) = \mua(\xvec)\left(H^{meas}(\xvec)-H(\xvec)\right)$, was independent of angle, photons were launched istropically with the launch position being spread out over the range of a source voxel using a randomly distributed number on the interval $[0,1]$. The initial photon weight was scaled according to the source strength with normalisation of the output quantity (i.e. radiance, absorbed energy density, harmonic, etc.) being $N_p$. The adjoint source may be negative in some places, so the initial photon weight is negative and weight deposition is also negative. The termination condition was therefore set to be the absolute value of the photon weight falling below the threshold value. The gradients were calculated using Eqs. \ref{grad_2D_abs_integral3} and \ref{scattering_gradient3}. A GD scheme was chosen for the minimisation because it is more robust to the MC noise in the functional gradients and error functional than techniques such as L-BFGS that use second-order information. A linesearch algorithm presented by Hager and Zhang \cite{Hager2005} was used for the reconstruction of $\mua$. A backtracking linesearch was implemented for the reconstruction of $\mus$. This is described in Section \ref{sec:mus-recon}.

The termination condition used by the optimisation was
\begin{equation}
\left|\epsilon^{(i)}-\epsilon^{(i-1)}\right| / \left|\epsilon^{(i)}\right|<10^{-9},
\label{termination}
\end{equation}
where $i$ is the iteration number.
For all the reconstructions it was assumed that the data $H^{meas}$ was given; no noise was added to the data, but MC noise from the forward simulation of the data was present at about 0.7\% (evaluated by taking several runs of the forward model to estimate the average standard deviation over all positions across all model runs). For each inversion, the forward and adjoint RMC simulations used 10\textsuperscript{8} photons and 10 Fourier harmonics, and was executed on a Dell 2U R820 32-core server.

\subsection{Inversion for absorption coefficient}\label{sec:mua-recon}
\begin{figure}[H]
\centering
\includegraphics[width=0.8\textwidth]{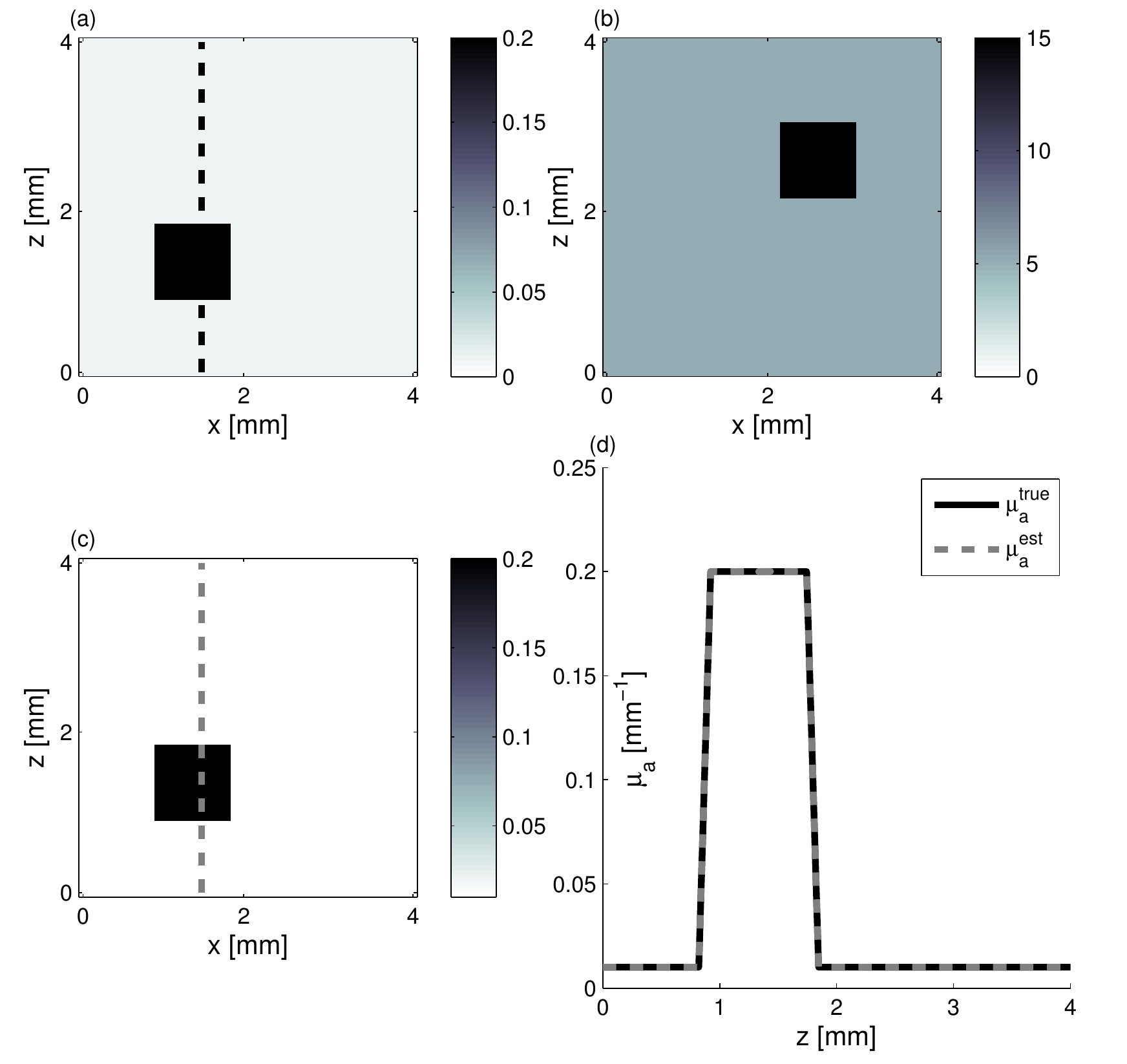}
\caption{(a) True absorption coefficient; (b) True scattering coefficient; (c) Reconstructed absorption coefficient after 9 iterations; (c) Profiles through true and reconstructed absorption coefficient at x=1.5mm for all z.}
\label{recon-mua}
\end{figure}

The domain used in the estimation of the absorption coefficient consisted of a background absorption coefficient of 0.01mm\textsuperscript{-1} with a rectangular inclusion equal to 0.2mm\textsuperscript{-1} (shown in Fig. \ref{recon-mua}(a)), and a background scattering coefficient of 5mm\textsuperscript{-1} with a rectangular inclusion equal to 15mm\textsuperscript{-1} (shown in Fig. \ref{recon-mua}(b)). The anisotropy was a homogeneously distributed value of 0.9. The measured data, $H^{meas}$, was formed using a MC simulation illuminated by a collimated line source on the boundary at z=0mm and on the adjacent boundary at x=4mm, consisting of 10\textsuperscript{8} photons. The inversion for the absorption coefficient only was performed under the assumption that the scattering coefficient was known and the starting estimate of the absorption coefficient was a homogeneous value of 0.01mm\textsuperscript{-1}. The termination condition in \eqref{termination} was satisfied after 11 iterations, having taken 4.1 hours to run, and is shown in Fig. \ref{recon-mua}(b) with profiles through the true and reconstructed distributions of $\mua$ shown in Fig. \ref{recon-mua}(d).

Very good agreement between the true and reconstructed absorption coefficient is observed, with a value of the error function after 11 iterations being 2.9$\times$10\textsuperscript{-9}. The optimisation routine was stopped as the change in the error function on the 12\textsuperscript{th} iteration was below the function tolerance of 10\textsuperscript{-9}, indicating convergence. The average error in the estimate of the absorption coefficient $\mua^{est}$, computed as $\left|\mua^{true}-\mua^{est}\right|/\mua^{true}$, was 0.2\% over the entire domain.

\subsection{Inversion for the scattering coefficient}\label{sec:mus-recon}
\begin{figure}[H]
\centering
\includegraphics[width=0.8\textwidth]{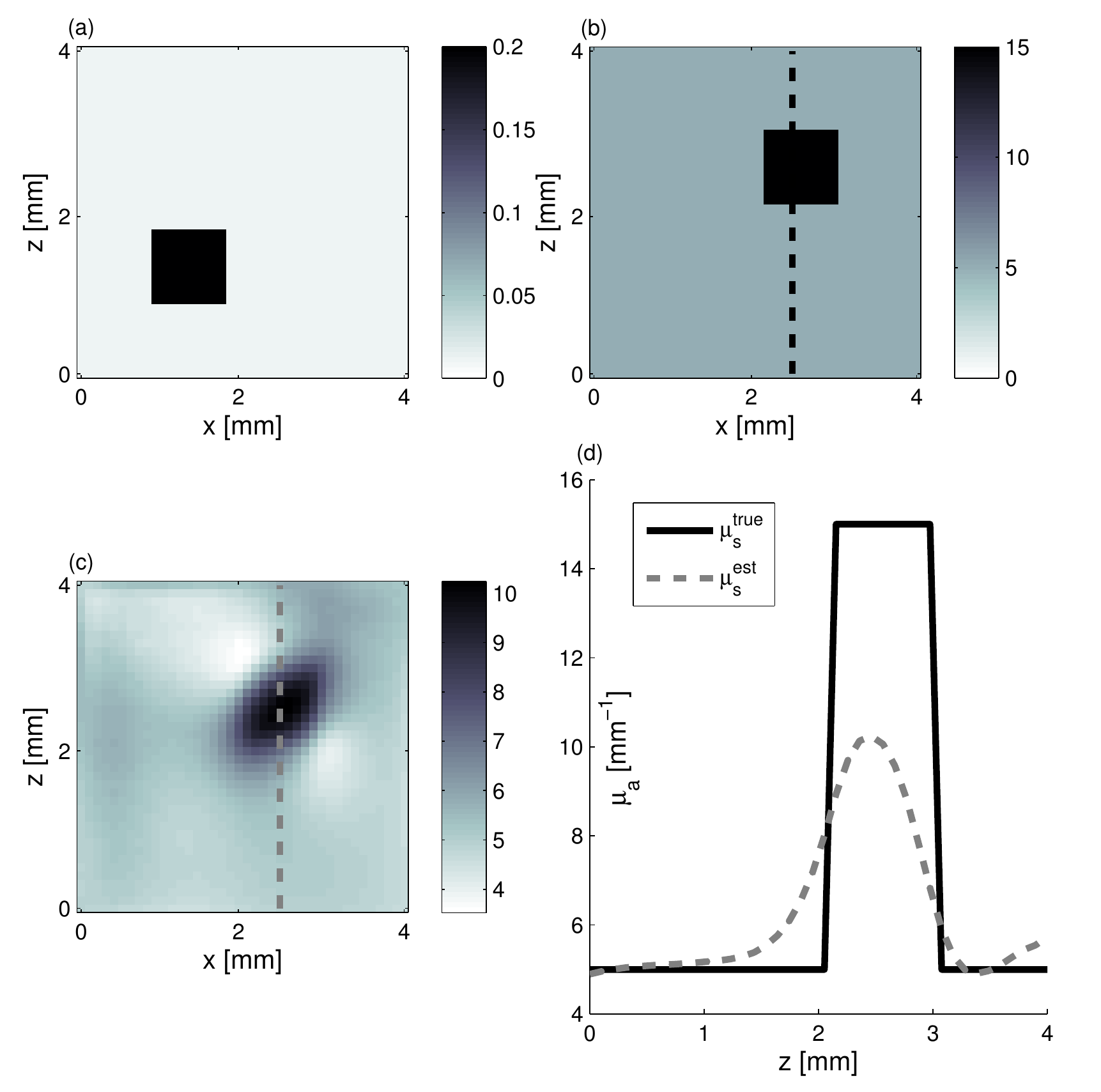}
\caption{(a) True absorption coefficient; (b) True scattering coefficient; (c) Reconstructed scattering coefficient after 35 iterations; (c) Profiles through true and reconstructed scattering coefficient at x=2.5mm for all z.}
\label{recon-mus}
\end{figure}

Inversions for the scattering coefficient were performed using the same domain as above, shown in Fig. \ref{recon-mus}(a) and (b), with the illumination and number of photons in the forward simulation also being the same. Here, it was assumed that the absorption coefficient was known and the starting estimate of the scattering coefficient was equal to the background value of 5mm\textsuperscript{-1}.

Two modifications were necessary to achieve convergence in the optimisation for the scattering coefficient. First, a custom GD algorithm was used in which a backtracking linesearch was implemented. A backtracking linesearch \cite{Nocedal1999} starts with a large candidate step length and progressively reduces the step size whilst checking for a sufficient decrease in the error functional. The sufficient decrease condition is expressed as
\begin{equation}\label{sufficient_decrease}
\epsilon([\mua^{(i)},\mus^{(i)}]+\alpha^{(i)}p^{(i)}) \leq \epsilon([\mua^{(i)},\mus^{(i)}])+\nu\alpha^{(i)}\nabla\epsilon^{(i)T}p^{(i)},
\end{equation}
where $\nu$ was chosen to be 0.2 by inspection because this produced rapid convergence. In order to improve efficiency of the linesearch, step sizes were bounded between $[10^{5},10^{9}]$; it was found that this range yielded sufficiently large steps to ensure reasonably efficient progress in the minimisation. Second, the termination condition in \eqref{termination} was relaxed due to the much slower convergence of the scattering coefficient, and instead required a relative change in the error functional of 10\textsuperscript{-5}. This was satisfied after 35 iterations, and is shown in Fig. \ref{recon-mus}(b) with profiles through the true and reconstructed distributions of $\mus$ shown in Fig. \ref{recon-mus}(d).

It can be seen from Fig. \ref{recon-mus}(c) and (d) that the inversion has partly reconstructed the inclusion in the scattering coefficient. The inability to reconstruct edges of the inclusion in the scattering coefficient is expected, given the diffusive nature of the scattering. However, the discrepancy in $\mus$ in the inclusion, evident from Fig. \ref{recon-mus}(d), suggests premature termination of the optimisation. This is due to the fact that the gradient with respect to scattering is small and prone to noise in the functional gradients. This low SNR in the gradients has the impact that that search directions in the optimisation routine are often sub-optimal, which results in little or no progress of the optimisation. The progressive reduction in SNR in the gradient means that non-descent steps are likely and can therefore trigger the termination condition.

\section{Discussion \& Conclusions}
\label{DC}
In this paper a novel MC model of the RTE was presented. The model computes the radiance in a Fourier basis in 2D and is straightforward to extend to 3D using a spherical harmonics basis. The accuracy of the model was demonstrated by comparing the angle-resolved radiance at two positions in the domain to corresponding appropriate analytic solutions.

Sections \ref{sec:gradients} and \ref{sec:examples} presented the application of the RMC algorithm to estimating the absorption and scattering coefficients from simulated PAT images. In Section \ref{sec:mua-recon} it was observed that the absorption coefficient was estimated with an average error of 0.2\% over the domain relative to the true value, when the scattering coefficient is known, and in the presence of 0.7\% average noise in the data. This is encouraging, particularly because noise is not only present in $H^{meas}$ but in the Fourier harmonics computed using the forward and adjoint RMC simulations, which is propagated to the estimates of the functional gradients. Consequently the search direction in the GD algorithm will always be sub-optimal. Furthermore, noise in $H(\mua^{(l)},\mus)$, the estimate of the absorbed energy density at the $l^{\text{th}}$ iteration of the linesearch, will be also be propagated to the error functional, resulting in a non-smooth search trajectory for the linesearch because at every point $\mua^{(l)}$, the error function will be corrupted by some different noise $\sigma^{(l)}$: $\epsilon(\mua^{(l)},\mus)=\left|\left|H^{meas}-H(\mua,\mus)(1+\sigma^{(l)})\right|\right|^{2}$. In practice, this did not preclude reconstruction of the absorption coefficient since the calculated gradients remained descent directions despite the noise. Furthermore, the error functional in $\mua$ is sufficiently convex that the addition of some noise does not prevent the linsearch from yielding sufficiently large a step length to allow rapid convergence.

Reconstruction of the scattering coefficent correctly located the scattering perturbation in the simulated image, however the peak value in the reconstruction was lower than the true value. This is a direct consequence of the fact that the scattering coefficient is related to the absorbed energy distribution only through the optical fluence distribution. Consequently, the SNR in $\frac{\partial\epsilon}{\partial\mus}$ is typically much less than that for absorption. This causes termination of the algorithm before the peak magnitude of the parameter has been found in the search space.

\section*{Acknowledgements}
The authors acknowledge the contribution of Andre Liemert who kindly provided radiance data used in the validation of RMC in Section \ref{sec:validation}.
The authors acknowledge the use of the UCL Legion High Performance Computing Facility (Legion@UCL), and associated support services, in the completion of this work. \\

\section*{Bibliography}
\bibliographystyle{spiebib}

\providecommand{\newblock}{}

\end{document}